\documentstyle[psfig,graphics]{mn2e}
\title [Galactic model parameters]
{A different approach for the estimation of Galactic model parameters
}
\author[Karaali et al.]
       {S.~Karaali,$^1$ \thanks{E-mail: karsa@istanbul.edu.tr}
        S.~Bilir$^1$, and E.~~Hamzao\u glu$^{2}$\\
  $^1$Istanbul University Science Faculty, Department of Astronomy and Space
      Sciences, 34119, University-Istanbul, Turkey\\
  $^2$Faculty of Engineering and Design, Istanbul Commerce University,  
      Rag$\i$p G\"um\"u\c spala Caddesi No: 84, 34378 
      Emin\"on\"u-Istanbul, Turkey\\}

\pagerange{\pageref{firstpage}--\pageref{lastpage}}

\begin{document}

\maketitle

\label{firstpage}

\begin{abstract}

We estimated the Galactic model parameters by means of a new approach based on 
the comparison of the observed space density functions per absolute magnitude 
interval with a unique density law for each population individually, and via 
the procedure in situ for the field SA 114 ($\alpha=22^{h}40^{m}00^{s},\delta= 
00^{o}00^{'}00{''}$; $l=68^{o}.15$, $b=-48^{o}.38$; 4.239 square-degree; epoch 2000). 
The separation of stars into different populations has been carried out by their 
spatial distribution. The new approach reveals that model parameters are absolute 
magnitude dependent. The scale height for thin disk decreases monotonously from 
absolutely bright ($M(g^{'})$=5) to absolutely faint ($M(g^{'})=13$) stars in a 
range 265-495 pc, but there is a discontunity at the absolute magnitude 
$M(g^{'})=10$ where the squared secans hiperbolicus density law replaces 
the exponential one. The range of the scale-height for thick disk, dominant in 
the absolute magnitude interval $5<M(g^{'})\leq9$, is less: 805-970 pc. The 
local space density for thick disk relative to thin disk decreases from 9.5\% 
to 5.2\% when one goes from the absolutely bright to faint magnitudes. Halo is 
dominant in three absolute magnitude intervals, i.e. $5<M(g')\leq6$, 
$6<M(g')\leq7$, and $7<M(g')\leq8$ and the axial ratio for this component is 
almost the same for these intervals where $c/a \sim 0.7$. The same holds for 
the local space density relative to the thin disk with range (0.02-0.15)\%. The 
model parameters estimated by comparison of the observed space density functions 
combined for three populations per absolute magnitude interval with the combined 
density laws agree with the cited values in the literature. Also each parameter 
is equal to at least one of the corresponding parameters estimated for different 
absolute magnitude intervals by the new approach. We argue that the most 
appropriate Galactic model parameters are those, that are magnitude dependent.  
             
\end{abstract}


\section{Introduction}

For some years, there has been a conflict among the researchers about the 
history of our Galaxy. Yet, there is a large improvement about this topic since 
the poineering work of Eggen, Lynden-Bell, \& Sandage\ (1962, hereafter ELS) who 
argued that the Galaxy collapsed in a free-fall time ($\sim 2\times 10^{8}$ yr). Now, 
we know that the Galaxy collapsed in times of many Gyr (e.g. Yoshii \& Saio\ 1979; 
Norris, Bessel, \& Pickles\ 1985; Norris 1986; Sandage \& Fouts\ 1987; Carney, 
Latham, \& Laird\ 1990; Norris \& Ryan\ 1991; Beers \& Sommer-Larsen\ 1995) and 
at least some of its components are formed from merger or accretion of numerous 
fragments, such as dwarf-type galaxies (cf. Searle \& Zinn\ 1978; Freeman \& 
Bland-Hawthorn\ 2002, and references therein). Also, the number of population 
components of the Galaxy increased from two to three, complicating 
interpretations of any dataset. The new component, 'thick disk', introduced 
by Gilmore \& Reid\ (1983) in order to explain the observation that star counts 
towards the South Galactic Pole were not in agreement with a single disk, 
'thin disk', component but rather could be nicely represented by two such 
components. The new component is discussed by Gilmore \& Wyse\ (1985) and 
Wyse \& Gilmore\ (1986).

The researchers use different methods to determine the parameters for three 
population components and  try to interpret them in relation to the formation 
and evolution of the Galaxy. Among the parameters, the local density and the 
scale-height of thick disk are the ones whose numerical values improved relative to the 
original ones claimed by Gilmore \& Reid\ (1983). In fact, the researchers  
indicate tendency to increase the original local density of thick disk from 2\% 
to 10\% relative to the total local density and to decrease its scale-height 
from the original value 1.45 kpc down to 0.65 kpc (Chen et al.\ 2001). In some 
studies, the range of the parameters is large especially for the thick disk. 
For example, Chen et al.\ (2001) and Siegel et al.\ (2002) give 6.5\% - 13\% and 
6\% - 10\% for the relative local density for thick disk, respectively. Now 
a question arises: what is the reason of these large ranges in two recent works 
(and in some other works) where one expects the most improved numerical 
values? This is the main topic of our paper. We argue that, though considerable 
improvements were achieved, we still couldn't chose the most appropriate 
procedure for the estimation of Galactic model parameters. In the present study, 
we show that, parameters of Galactic model are functions of absolute magnitude 
and if this is not taken into consideration, different parameters with large 
ranges are obtained and this can not be unavoided. 

In sections 2 and 3 different methods and density law forms are discussed. 
The procedure used in this study is given in section 4. Section 5 provides 
de-reddened apparent magnitude, absolute magnitude, distance and density 
function determinations. Model parameter estimation by different procedures 
is given in section 6 and finally section 7 provides discussion.

\section{The methods}

The studies related to the Galactic structure are usually carried out by 
star counts. Direct comparison between the theoretical and observed space 
densities is also used as another method. The first method is based 
on the fundamental equation of stellar statistics (von Seelinger\ 1898) which 
may be written as follows:

\begin{eqnarray}
A(m_{V},S_{B-V}) = \sum A_{i}(m_{v},S_{B-V})\nonumber\\
= \Omega \sum \int \Phi_{i}(M,S)D_{i}(r)r^{2}dr 
\end{eqnarray}
where $A$ is the differential number of counts at any particular magnitude 
and colour, $A_{i}$ is the contribution to those counts from population $i$,
$\Omega$ is the solid angle observed, $\Phi_{i}$ is the luminosity function 
of population $i$, and $D_{i}$ is the density distribution of population $i$ 
as a function of absolute magnitude $M$ and spectral type $S$, and 
$r$ is distance along the line of sight. Here, the number of counts is the 
sum over the stellar populations of the convolution of the luminosity and 
density distribution functions. It is stated by many authors (cf. Siegel et 
al.\ 2002), the noninvertability and the vagaries of solving the nonunique 
convolution by trial and error, limit the star counts and will be a weak tool for 
exploring the Galaxy. The large number of Galactic structure models derived 
from star count studies confirms the nonuniquness problem (Table 1). The most 
conspicuous point in Table 1 is the large range of thick disk parameters, 
indicating a less certain density law for this component. Whereas the thin 
disk parameters occupy a narrow range of values. For halo, the results 
from star count surveys cover almost the entire range of parameter space 
from flattened de Vaucouleurs spheroid (Wyse \& Gilmore\ 1989, Larsen\ 1996) 
to perfectly spherical power-law distributions (Ng et al.\ 1997). 

There is not enough study in the literature, carried out by comparison of theoretical 
and observational space densities. The works of Basle group (del Rio \& 
Fenkart\ 1987, and Fenkart \& Karaali\ 1987), and recently, Phleps et al.\ (2000), 
Siegel et al.\ (2002), Karaali et al.\ (2003), and Du et al.\ (2003) can be mentioned 
as examples. Photometric parallaxes provide direct evaluation of spatial 
densities. Hence, the observations can be translated into discrete density 
measurements at various points in the Galaxy, instead of trying to fit the 
structure of the Galaxy into the observed parameter space of colours and 
magnitudes.

Almost the same results are seen in several studies. This is not a surprise, 
since such studies have explored similar data sets with similar limitations and 
additionally they probe the same direction in the sky such as the Galactic poles. 
Most studies are based on investigation  of one or a few fields in different 
directions. The deep fields are small with corresponding poor statistical weight 
and the large fields are limited with shallower depth which may not be able to 
probe the Galaxy at large distances. The works of Reid \& Majewski\ (1993) and 
Reid et al.\ (1996) can be given as examples for the first category, while the 
one of Gilmore \& Reid\ (1983) for the second category. Deep surveys based on 
the multidirectional Hubble Space Telescope (HST) (Zheng et al.\ 2001), 
has another limitation, i.e. star-galaxy separation becomes difficult at faint 
magnitudes. There are few programs which survey the Galaxy in multiple 
directions such as the Basle Halo Program (cf. Buser, Rong, \& Karaali\ 1999), 
the Besan\c{c}on program (cf. Robin et al.\ 1996, 2003), the APS-POSS program 
(Larsen\ 1996), and recently the SDSS (Chen et al.\ 2001). 

Star count studies in a single direction can lead to degenerate solutions and 
surveys, limited with small areas at the Galactic poles are insensitive to radial 
terms in the population distributions (Reid \& Majewski\ 1993, Robin et al.\ 1996, 
and Siegel et al.\ 2002). 

\begin{table*}
\center
\caption{Previous Galactic models. Symbols: TN: thin disk, TK: thick disk,
S: spheroid (halo), $R_{e}$: effective radius, $c/a$: axes ratio. The figures 
in the parentheses for Siegel et al.\ (2002) are the corrected values 
for binarism. (*) power-law index replacing $R_{e}$.}
{\scriptsize
\begin{tabular}{lllllllll}
\hline
H (TN)& h(TN)& n (TK)& H(TK) & h(TK) & n (S) & $R_{e}$(S) & c/a &  Reference \\
(pc) & (kpc) & &  (kpc) & (kpc) & & (kpc) & & \\
\hline
310-325 & --- & 0.0125-0.025 &  1.92-2.39 & --- & --- & --- & --- & Yoshii\ 1982 \\
300 & --- & 0.02 & 1.45 & --- & 0.0020 & 3.0 & 0.85 & Gilmore \& Reid\ 1983\\
325 & --- & 0.02 & 1.3 &  --- & 0.0020 & 3.0 & 0.85 & Gilmore\ 1984 \\
280 & --- & 0.0028 &  1.9 & --- &0.0012 &--- & --- & Tritton \& Morton\ 1984 \\
200-475 & --- & 0.016 &  1.18-2.21 & --- & 0.0016 & --- & 0.80 & 
Robin \& Cr\'{e}z\'{e}\ 1986 \\
300 & --- & 0.02 & 1.0 & --- & 0.0010 & --- & 0.85 & del Rio \& Fenkart\ 1987 \\
285 & --- & 0.015 &1.3-1.5 & --- & 0.0020 & 2.36 & Flat & Fenkart et al.\ 1987 \\
325 & --- & 0.0224 & 0.95 & --- & 0.0010 & 2.9 & 0.90 & Yoshii et al.\ 1987 \\
249 & --- & 0.041 & 1.0 & --- & 0.0020 & 3.0 & 0.85 & Kuijken \& Gilmore\ 1989 \\
350 & 3.8 & 0.019 & 0.9 & 3.8 & 0.0011 & 2.7 & 0.84 & Yamagata \& Yoshii\ 1992 \\
290 & --- & --- & 0.86 & --- & --- & 4.0 & --- & VonHippel \& Bothun\ 1992 \\
325 & --- & 0.0225 & 1.5 & --- & 0.0015 & 3.5 & 0.80 & Reid \& Majewski\ 1993 \\
325 & 3.2 & 0.019 & 0.98 & 4.3 & 0.0024 & 3.3 & 0.48 & Larsen\ 1996 \\
250-270 & 2.5 & 0.056 & 0.76 & 2.8 & 0.0015 & 2.44-2.75* &  0.60-0.85 & 
Robin et al.\ 1996, 2000 \\
290 &4.0 & 0.059 & 0.91 & 3.0 & 0.0005 & 2.69 & 0.84 & Buser et al.\ 1998, 1999 \\
240 &2.5 & 0.061 & 0.79 & 2.8 & --- & --- & 0.60-0.85 & Ojha et al.\ 1999 \\
330 &2.25& 0.065-0.13 &  0.58-0.75 & 3.5 & 0.0013 & --- & 0.55 & 
Chen et al.\ 2001 \\
280(350) & 2-2.5&  0.06-0.10 & 0.7-1.0 (0.9-1.2) & 3-4 & 0.0015 & --- & 
0.50-0.70 & Siegel et al.\ 2002\\
\hline
\end{tabular}
}  
\end{table*}

\section{The density law forms}

Disk structures are usually parameterised in cylindrical coordinates 
by radial and vertical exponentials,

\begin{eqnarray}
\tiny
D_{i}(x,z)=n_{i}e^{-z/H_{i}}e^{-(x-R_{o})/h_{i}}
\end{eqnarray}
where $z$ is the distance from Galactic plane, $x$ is the planar distance 
from the Galactic center, $R_{0}$ is the solar distance to the Galactic 
center (8.6 kpc), $H_{i}$ and $h_{i}$ are the scale height and scale length 
respectively, and $n_{i}$ is the normalized local density. The suffix $i$ 
takes the values 1 and 2, as long as thin and thick disks are considered. 
A similar form uses the $sech^{2}$ (or $sech$) function to parameterize 
the vertical distribution for thin disk,

\begin{eqnarray}
D_{i}(x,z)=n_{i}sech^{2}(z/H^{'}_{i})e^{-(x-R_{o})/h_{i}}.
\end{eqnarray}
As the secans hyperbolicus is the sum of two exponentials, $H^{'}_{i}$ is 
not really a scale height, but has to be compared to $H_{i}$ by multiplying 
it with $arcsech(1/e)\sim 1.65745: H_{i}=1.65745H^{'}_{i}$.

The density law for spheroid component is parameterised in different forms. 
The most common is the de Vaucouleurs\ (1948) spheroid used to describe the 
surface brightness profile of elliptical galaxies. This law has been 
deprojected into three dimensions by Young\ (1976) as follows:

\begin{eqnarray}
D_{s}(R)=n_{s}~exp[-7.669(R/R_{e})^{1/4}]/(R/R_{e})^{0.875}
\end{eqnarray}
where $R$ is the (uncorrected) Galactocentric distance in spherical 
coordinates, $R_{e}$ is the effective radius, and $n_{s}$ is the normalised 
local density. $R$ has to be corrected for the axial ratio $\kappa = c/a$, 

\begin{eqnarray}
R = [x^{2}+(z/\kappa)^2]^{1/2}
\end{eqnarray}
where,
\begin{eqnarray}
x = [R_{o}^{2}+(z/\tan b)^2-2R_{o}(z/\tan b)\cos l]^2
\end{eqnarray}
$b$ and $l$ being the Galactic latitude and longitude for the field under 
investigation. The form used by the Basle group is independent of effective 
radius but the distance of the Sun to the Galactic center: 

\begin{eqnarray}
D_{s}(R)=n_{s}~exp[10.093(1-R/R_{o})^{1/4}]/(R/R_{o})^{0.875}
\end{eqnarray}
and alternative formulation is the power-law,
\begin{eqnarray}
D_{s}(R)=n_{s}/(a_{o}^{n}+R^{n})
\end{eqnarray}
where $a_{o}$ is the core radius.
	         
Equations (2) and (3) replaced by (9) and (10) respectively, as long as 
vertical direction is considered, where:

\begin{eqnarray}
D_{i}(z)=n_{i}e^{-z/H_{i}}
\end{eqnarray}

\begin{eqnarray}
D_{z}(z)=n_{i}sech^{2}(-z/H_{i}^{'})
\end{eqnarray}

\section{The procedure used in this work}

In this work, we compared the observed and the theoretical space densities per 
absolute magnitude interval in the vertical direction of the Galaxy for a large 
range of absolute magnitude interval, $4<M(g^{'})\leq13$, down to the limiting 
magnitude $g^{'}=22$. The procedure is similar to that of Phleps et al.\ (2000), 
however the approach is different. First, we separated the stars into different 
populations by a slight modification of the method given by Karaali\ (1994), i.e. 
we used the spatial distribution of stars as a function of both absolute magnitude 
and apparent magnitude, whereas Karaali prefered a unique distribution for stars 
of all apparent magnitudes. Second, we derived model parameters for each population 
individually for each absolute magnitude interval and we observed their difference. 
Third, and finally, the model parameters are estimated by comparison of the 
observed vertical space densities with the combined density laws (equations 7, 9, 
and 10) for stars of all populations. In the last process we obtained two sets of 
parameters, one for the absolute magnitude interval $5<M(g^{'})\leq10$, and another 
one for $5<M(g^{'})\leq13$. We notice that the behavior of stars with 
$10<M(g^{'})\leq13$ is different. We must keep in mind that many researchers related to 
estimation of the model parameters are restricted with $M(V)\leq8$ (cf. Robin et al.\ 
2003). Different behavior of stars with $10<M(g^{'})\leq13$ may result in different 
values and large ranges may be expected for the model parameters based on star counts.

\section{The data and reductions}

\subsection{Observations}
Field SA 114 ($\alpha=22^{h}40^{m}00^{s},\delta=00^{o}00^{'}00{''}$; 
$l=68^{o}.15$, $b=-48^{o}.38$; 4.239 square-degree; epoch 2000) was measured by the 
Isaac Newton Telescope (INT) Wide Field Camera (WFC) mounted at the prime focus 
($f/3$) of the 2.5m INT on La Palma, Canary Islands, during seven observing runs, 
i.e. 1998 September 3, 1999 July 17-22, 1999 August 18, 1999 October 9-10, 2000 
June 25-30, 2000 October 19, and 2000 November 15-22. The WFC consists of 4 EVV42 
CCDs, each containing 2k x 4k pixels. They are fitted in a L-shaped pattern which 
makes the camera 6k x 6k, minus 2k x 2k corner. The WFC has 13.5$\mu$ pixels 
corresponding to 0.33 arcsec pixel$^{-1}$ at INT prime focus and each covers an 
area of 22.8 x 11.4 arcmin on the sky. A total of 0.29 square-degree is covered by 
the combined of four CCDs. With a typical seeing of 1.0-1.3 arcsec on the INT, 
point objects are well sampled, which allows accurate photometry.  

Observations were taken in five bands ($u^{'}_{RGO}$, $g^{'}$, $r^{'}$, $i^{'}$, 
$z^{'}_{RGO}$) with a single exposure of 600s to nominal 5$\sigma$ limiting 
magnitudes of 23, 25, 24, 23, and 22, respectively (McMahon et al.\ 2001). 
Magnitudes are put on a standart scale using observations of Landolt standard 
star fields taken on the same night. The accuracy of the preliminary 
photometric calibration is $\pm$0.1 mag.     

\begin{figure}
\resizebox{8cm}{5.2cm}{\includegraphics*{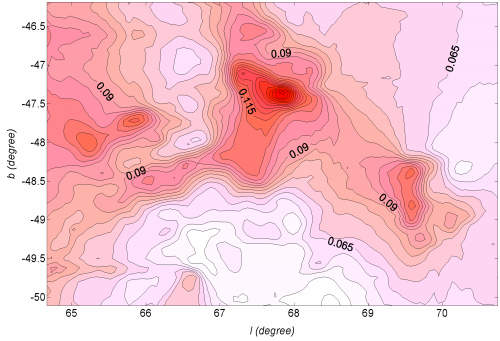}} 
\caption[] {$E(B-V)$ colour-excess contours for the field SA 114 as a function 
of Galactic latitude and longitude.}
\end {figure}

\begin{figure}
\resizebox{8cm}{5cm}{\includegraphics*{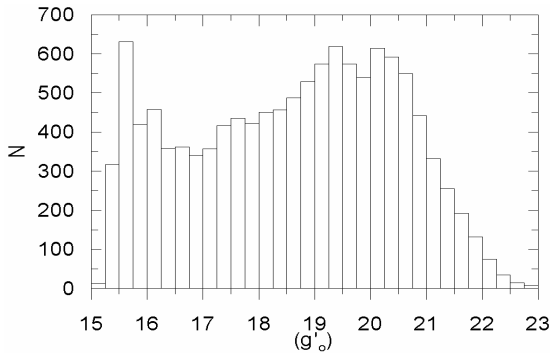}} 
\caption[] {Apparent magnitude histogram for all sources. Bright stars are 
effected by saturation.}
\end{figure}

\subsection{The overlapping sources, de-reddening of the magnitudes, bright stars, 
and extra-galactic objects}

The magnitudes are provided from the Cambridge Astronomical Survey Unit (CASU). 
There are totally 14439 sources in 24 sub-fields in the field SA 114. It turned 
out that 2428 of these sources are overlapped, i.e. their angular distances are 
less than $1{''}$ to any other source. We omitted them, thus the sample reduced 
to 12011. The $E(B-V)$ colour-excesses for the sample sources are evaluated by 
the procedure of Schegel, Finkbeiner, \& Davis,\ (1998) and corrected by the 
following equations of Beers et al.\ (2002):

\begin{figure}
\resizebox{6cm}{11.2cm}{\includegraphics*{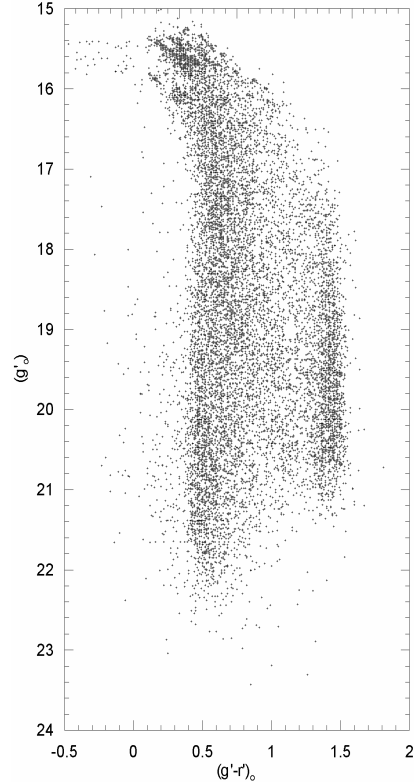}} 
\caption[] {Colour-apparent magnitude diagram for the original sample.}
\end{figure}

\begin{eqnarray}
E(B-V)=E(B-V)_{s},~~~~for~(B-V)s\leq0.10\\  
E(B-V)=0.10+0.65[E(B-V)_{s}-0.10 ], \nonumber\\
~~~~for~E(B-V)s > 0.10
\end{eqnarray}
where $E(B-V)_{s}$ is the colour-excess evaluated via the procedure of Schlegel 
et al.\ The $E(B-V)$ colour-excess contours for the field is given in Fig. 1 as a 
function of Galactic latitude and longitude. Then, the total absorption $A_{V}$ 
is evaluated by means of the well known equation,

\begin{eqnarray}
R_{V}=\frac {A_{V}} {E(B-V)} = 3.1					
\end{eqnarray}

For Vega bands we used the $R_{\lambda}/R_{V}$ data of Cox (2000) for the 
interpolation, where $\lambda$ = 3581$\AA$, 4846$\AA$, 6240$\AA$, 7743$\AA$, and 
8763$\AA$ (Table 2), and derived $R_{\lambda}$ by their combination with $A_{V}$. 
Finally, the de-reddened $u_{0}^{'}$, $g_{0}^{'}$, $r_{0}^{'}$, $i_{0}^{'}$, and 
$z_{0}^{'}$ magnitudes were obtained from the original magnitudes and the 
corresponding $R_{\lambda}$.

\begin{figure}
\center
\resizebox{5cm}{10.6cm}{\includegraphics*{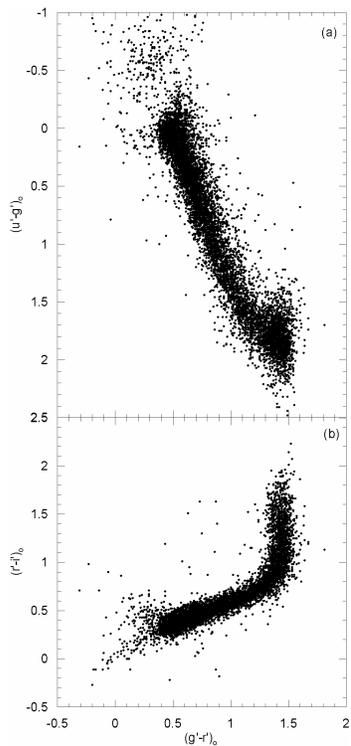}} 
\caption[] {Two-colour diagrams for \it sources  \rm with apparent magnitude 
$17<g^{'}_{0}\leq22$. (a) for $(u^{'}-g^{'})_{0}-(g^{'}-r^{'})_{0}$ and 
(b) for ($g^{'}-r^{'})_{0}-(r^{'}-i^{'})_{0}$.}
\end{figure}

\begin{figure}
\center
\resizebox{5cm}{10.6cm}{\includegraphics*{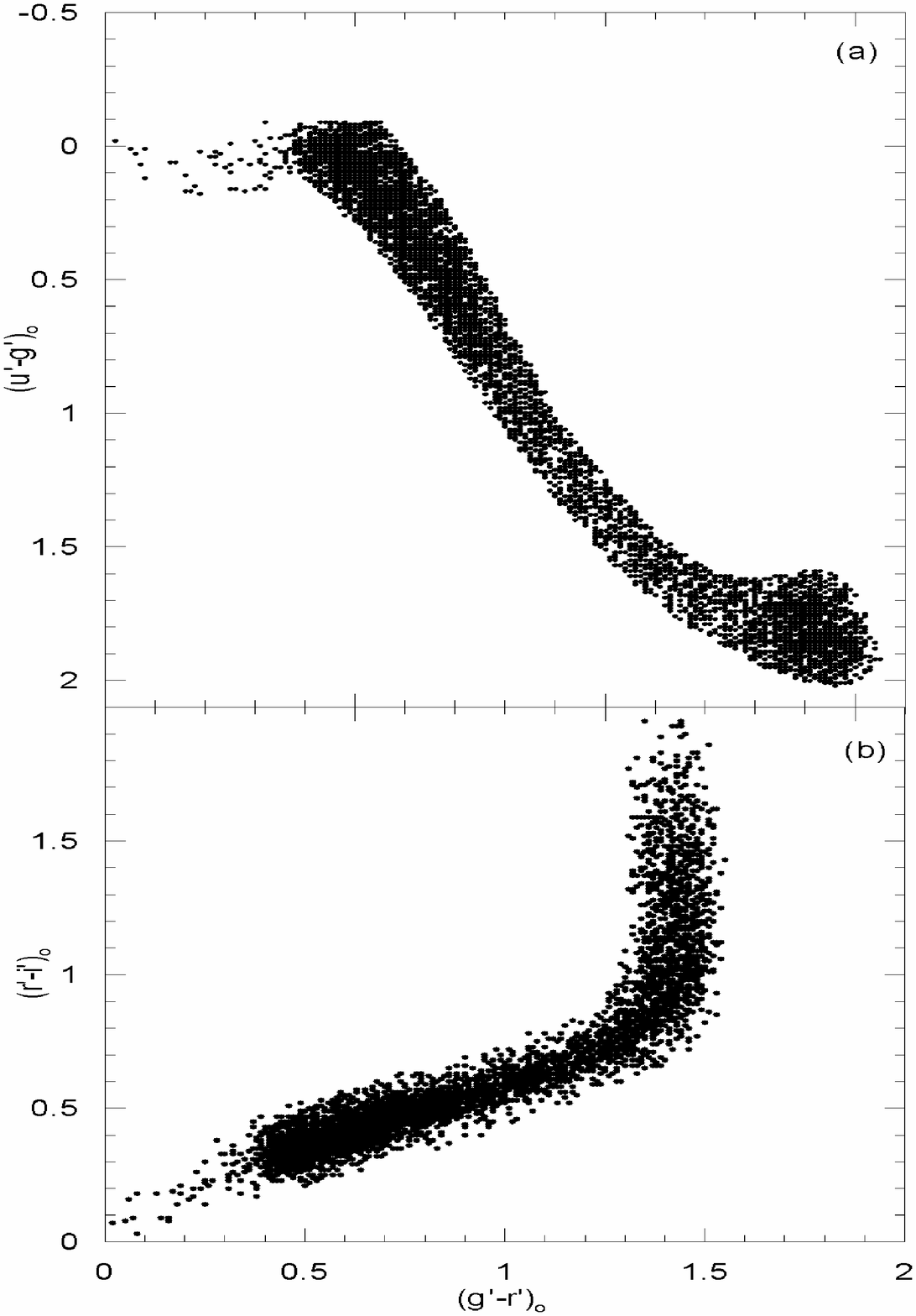}} 
\caption[] {Two-colour diagrams for \it stars \rm with apparent magnitude. 
$17<g^{'}_{0}\leq22$. (a) for $(u^{'}-g^{'})_{0}-(g^{'}-r^{'})_{0}$ and 
(b) for ($g^{'}-r^{'})_{0}-(r^{'}-i^{'})_{0}$.}
\end{figure}

\begin{table}
\center
\caption{The relation between the total absorptions for Vega bands and for $V$ band 
of $UBV$ photometry.}
\begin{tabular}{ccccc}
\hline
Filter & $\lambda_{eff}(\AA)$ & $\Delta_{eff}(\AA)$ & $R_{i}/R_{v}$ & $m_{lim}$ \\
\hline
$u^{'}$ & 3581 &638 & 1.575& 24.3 \\
$g^{'}$ & 4846 &1285& 1.164& 25.2 \\
$r^{'}$ & 6240 &1347& 0.867& 24.5 \\
$i^{'}$ & 7743 &1519& 0.648& 23.7 \\
$z^{'}$ & 8763 &950 & 0.512& 22.1 \\
\hline
\end{tabular}  
\end{table}

The histogram for the de-reddened apparent magnitude $g_{0}^{'}$ (Fig. 2) 
and the colour-apparent magnitude diagram (Fig. 3) show that there is a large 
number of saturated sources in our sample. Hence, we excluded sources brighter 
than $g_{0}^{'}=17$. However, the two-colour diagrams $(u^{'}-g^{'})_{0}$ - 
$(g^{'}-r^{'})_{0}$ and $(g^{'}-r^{'})_{0}$ - $(r^{'}-i^{'})_{0}$ in Fig. 4 
indicate that there are also some extra-galactic objects, where 
most of them lie towards the blue as claimed by Chen et al. (2001). It seems 
that the star/extragalactic object separation based on the 'stellarity 
parameter' as returned from the SExtractor routines (Bertin \& Arnouts,\ 1996) 
couldn't be sufficient. This parameter has a value between 0 (highly extended), 
and 1 (point source). The separation work very well to classify a point source 
with a value greater than 0.8. We adopted the simulations of Fan (1999), 
additional to the work cited above to remove the extragalactic objects in our 
field. Thus we rejected  the sources with $(u^{'}-g^{'})_{0}<-0.10$, and those 
which lie outside of the band concentrated by most of the sources. After the last 
process the number of sources in the sample, stars, reduced to 6418. The 
two-colour diagrams $(u^{'}-g^{'})_{0}$ - $(g^{'}-r^{'})_{0}$ and 
$(g^{'}-r^{'})_{0}$ - $(r^{'}-i^{'})_{0}$ for the final sample are given in 
Fig. 5. A few dozen of stars with $(u^{'}-g^{'})_{0}\sim0.10$ and 
$(g^{'}-r^{'})_{0}\sim0.20$ are probably stars of spectral type A.

\begin{figure}
\resizebox{5.2cm}{11.20cm}{\includegraphics*{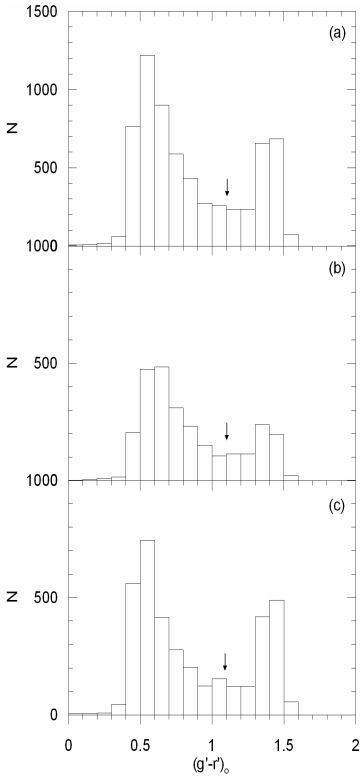}} 
\caption[] {$(g^{'}-r^{'})_{0}$ colour histogram as a function of apparent 
magnitude, for the star sample. (a) for $17<g^{'}_{0}\leq22$, (b) for 
$17<g^{'}_{0}\leq19$, and (c) for $19<g^{'}_{0}\leq22$. The vertical 
downward arrow shows the limit value $(g^{'}-r^{'})_{0}=1.1$ mag which 
separates thin disk and thick disk - halo couple.}
\end{figure}

\begin{figure*}
\resizebox{16cm}{8.34cm}{\includegraphics*{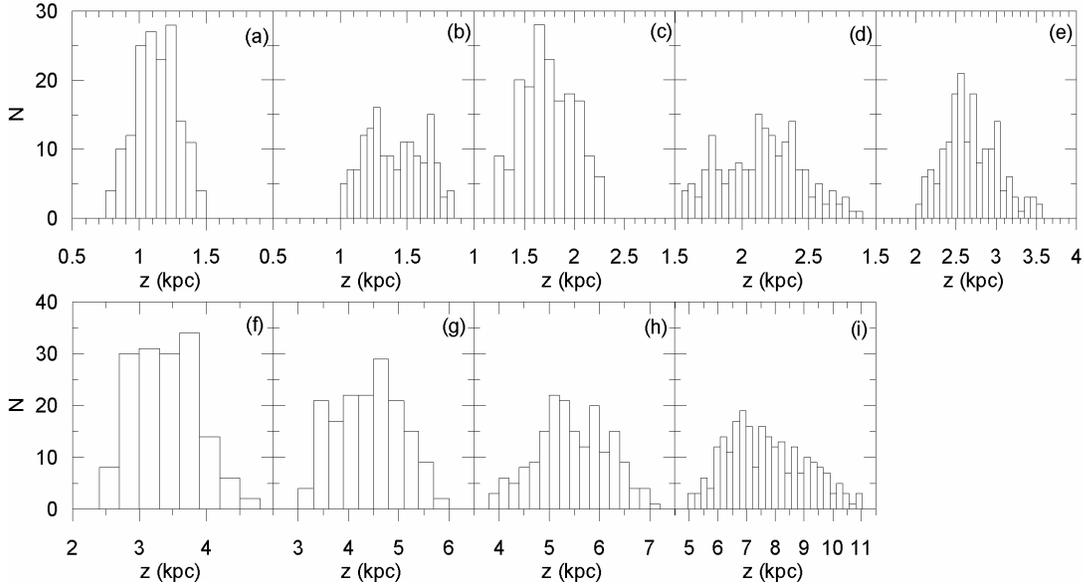}} 
\caption[] {Spatial distribution for stars with absolute magnitude 
$6<M(g^{'})\leq7$ as a function of apparent magnitude. (a)(17.0-17.5], 
(b) (17.5-18.0], (c)(18.0-18.5], (d)(18.5-19.0], (e)(19.0-19.5], 
(f)(19.5-20.0], (g) (20.0-20.5], (h)(20.5-21.0], and (i) (21-22]}
\end{figure*}

\subsection{Absolute magnitudes, distances, population types and density functions}

In the $Sloan$ photometry, the blue stars in the range $15<g^{'}<18$ are dominated 
by thick disk stars with a turnoff $(g^{'}-r^{'})\sim0.33$, and for $g^{'}>18$, the 
Galactic halo, has a turnoff colour $(g^{'}-r^{'})\sim0.2$, becomes significant. 
Red stars, $(g^{'}-r^{'})\geq1.3$, are dominated by thin disk stars for all apparent 
magnitudes (Chen et al.\ 2001). In our case, the apparent magnitude which separates 
the thick disk and halo stars seems to be a bit fainter relative to the $Sloan$ 
photometry, i.e. $g_{0}^{'}\sim19$ (Fig. 3). Thus, stars bluer than 
$(g^{'}-r^{'})=1.1$ and brighter than $g_{0}^{'}=19$ are separated to the thick 
disk population and the colour-magnitude diagram of 47 Tuc (Hesser et al.\ 1987) is 
used for their absolute magnitude determination, whereas those with the same colour 
but fainter than $g_{0}^{'}=19$ are assumed to be as halo stars and their absolute 
magnitudes are determined via the colour-magnitude diagram of M92 (Stetson \& 
Harris\ 1988). From the other hand, stars redder than $(g^{'}-r^{'})=1.1$ are 
adopted as thin disk stars and their absolute magnitudes are evaluated by means 
of the colour-magnitude diagram of Lang\ (1992) for Pop I stars (Fig. 6). The 
distance to a star relative to the Sun is carried out by the following formula:

\begin{eqnarray}
(g^{'}-M(g^{'}))_{o} =  5 \log r - 5					
\end{eqnarray}
The vertical distance to the galactic plane ($z$) of a star could be evaluated by 
its distance $r$ and its Galactic latitude ($b$) which could be provided by its 
right ascension and declination. The precise separation of stars into different 
populations have been carried out by their spatial distribution as a function of 
their absolute and apparent magnitudes. Fig. 7 gives the spatial distribution of 
stars with $6<M(g^{'})\leq7$, as an example, and Table 3 gives the full set of 
absolute and apparent magnitude intervals, and the efficiency regions of the 
populations. Halo stars dominate the absolutely bright intervals, thick disk 
stars indicate the intermediate and the thin disk stars the faint ones, as 
expected (Fig. 8).

Number of stars as a function of distance $r$ relative to the Sun, and the 
corresponding mean distance $z^{*}$ from the Galactic plane, for different absolute 
magnitude intervals for three populations are given in Tables 4-6. The logarithmic 
space density functions, $D^{*}=\log D+10$, evaluated by means of these data are 
not given in the tables for avoidance space consuming but they are presented in Fig. 
9-11, where $D=N/ \Delta V_{1,2}$, $\Delta V_{1,2}=(\pi/180)^{2}(\sq/3)
(r_{2}^{3}-r_{1}^{3})$, $\sq$: size of the field (4.239 square-degree), $r_{1}$ and 
$r_{2}$: the limiting distance of the volume $\Delta V_{1,2}$, and N: number of stars 
(per unit absolute magnitude). The horizontal thick lines, in Tables 4-6, 
corresponding to the limiting distance of completeness ($z_{l}$) are evaluated by 
the following equations:

\begin{table}
\center
\caption{Dominant regions for three populations; thin disk, thick disk, 
and halo, as a function of absolute and apparent magnitudes. The symbol 
$z$ is the distance to the Galactic plane.}
{\scriptsize
\begin{tabular}{ccccc}
\hline
{$M(g^{'})$}&{$g_{o}^{'}$}&{Thin disk}&{Thick disk}&{Halo}\\
\hline
(12-13] & (17-22]    &   $z\leq0.50$ &$---$    & $---$ \\
(11-12] & (17-22]    &   $z\leq0.80$ &$---$    & $---$ \\
(10-11] & (17-22]    &   $z\leq1.30$ &$---$    & $---$ \\
(9-10]	& (17-22]    &   $z\leq1.50$ &$z>1.50$ & $---$ \\
(8-9]	& (17-18]    &   $z\leq0.60$ &$z>0.60$ & $---$ \\
        & (18-19]    &   $z\leq0.85$ &$z>0.85$ & $---$ \\
	& (19-20]    &   $z\leq1.25$ &$z>1.25$ & $---$ \\
	& (20-22]    &   $z\leq1.60$ &$z>1.60$ & $---$ \\
(7-8]	& (17-18]    &   $z\leq1.00$ &$z>1.00$ & $---$ \\
	& (18-19]    &   $z\leq1.30$ &$z>1.30$ & $---$ \\
	& (19.0-19.5]&$z\leq1.70$    &$z>1.70$ & $---$ \\
	& (19.5-20.0]&$---$         &$z\leq2.55$  & $z>2.55$ \\
	& (20-22]    &$---$         &$z\leq3.80$  & $z>3.80$ \\
(6-7] & (17.0-17.5]&$z\leq1.12$   &$z>1.12$& $---$ \\
	& (17.5-18.0]&$z\leq1.40$   &$z>1.40$& $---$ \\
	& (18.0-18.5]&$z\leq1.50$   &$z>1.50$& $---$ \\
	& (18.5-19.0]&$z\leq1.90$   &$z>1.90$& $---$ \\
	& (19.0-19.5]&$z\leq2.25$   &$z>2.25$& $---$ \\
	& (19.5-20.0]&$---$         &$z\leq4.60$  & $---$ \\
	& (20.0-20.5]&$---$         &$z\leq4.70$  & $z>4.70$ \\
	& (20.5-21.0]&$---$         &$z\leq5.60$  & $z>5.60$ \\
	& (21-22]    &$---$         &$z\leq6.40$  & $z>6.40$ \\
(5-6]   & (17.0-18.0]&$z\leq1.68$   &$z>1.68$& $---$ \\
	& (18.0-18.5]&$z\leq2.18$   &$z>2.18$& $---$ \\
	& (18.5-19.0]&$z\leq2.52$   &$z>2.52$& $---$ \\
	& (19.0-19.5]&$---$         &$z\leq4.50$   & $z>4.50$ \\
	& (19.5-20.0]&$---$         &$z\leq4.50$   & $z>4.50$ \\
	& (20-22] &  $---$          &$---$         & $z>4.50$ \\
(4-5]   & (17-18] &  $---$          &$z\leq2.20$   & $z>2.20$ \\
	& (18-19] &  $---$          &$z\leq3.60$   & $z>3.60$ \\
	& (19-22] &  $---$          &$---$         & $z>5.00$ \\
\hline
\end{tabular}
}
\end{table}

\begin{table*}
\center
\caption{Number of stars as a function of distance $r$ relative to the Sun, and 
the corresponding mean distance $z*$ from the Galactic plane, for different 
absolute magnitude intervals for thin disk (distances in kpc). Horizontal thick 
lines correspond the limiting distance of completeness.}
{\scriptsize
\begin{tabular}{lrrrrrrrrrrrrrrrr}
\hline
$M(g^{'}) \rightarrow$ & \multicolumn{2} {c} {(5-6]} &\multicolumn{2} {c} {(6-7]} & \multicolumn{2} {c} {(7-8]} & \multicolumn{2} {c} {(8-9]} & \multicolumn{2} {c} {(9-10]} & \multicolumn{2} {c} {(10-11]} & \multicolumn{2} {c} {(11-12]} & \multicolumn{2} {c} {(12-13]} \\
\hline
$r_{1}-r_{2}$ & $z^{*}$ & N &$z^{*}$ & N & $z^{*}$ & N &$z^{*}$& N &$z^{*}$ & N &$z^{*}$& N & $z^{*}$&  N &$z^{*}$& N \\
\hline
0.10-0.20 & & & & & & & & & & & & & 0.13 & 30 & 0.12 & 20 \\ \cline{14-15}
0.20-0.30 & & & & & & & & & 0.22 & 4 & 0.19 & 42 & 0.19 & 115 & 0.19 & 39 \\ \cline{12-13}
0.30-0.40 & & & & & & & & & 0.27 & 21& 0.26 & 83 & 0.26 & 129 & 0.26 & 33 \\ \cline{10-11} \cline{16-17}
0.40-0.60 & & & & & & & 0.43 & 59 & 0.37 & 66 & 0.37 & 154 & 0.36 & 182 & 0.34 & 28 \\ \cline{8-9}
0.60-0.80 & & & & & 0.56 & 14 & 0.52 & 65 & 0.52 & 58 & 0.52 & 158 & 0.50 & 71 &  & \\ \cline{14-15}
0.80-1.00 & & & & & 0.67 & 43 & 0.68 & 58 & 0.68 & 72 & 0.67 &  87 & 0.66 & 20 &  & \\ \cline{6-7}
1.00-1.25 & & & 0.87 & 18 & 0.84 & 87 & 0.82 & 65 & 0.85 & 72 & 0.83 & 77 & & & &  \\ \cline{4-5} \cline{12-13}
1.25-1.50 & & & 1.03 & 67 & 1.02 & 76 & 1.03 & 55 & 1.03 & 46 & 1.01 & 30 & & & &  \\ \cline{10-11}
1.50-1.75 & 1.26 & 8 & 1.21 & 60 & 1.21 & 59 & 1.20 & 66 & 1.20 & 37 & 1.24 & 7 & & & & \\
1.75-2.00 & 1.42 &39 & 1.39 & 46 & 1.40 & 37 & 1.40 & 16 & 1.40 & 16 & & & & & & \\ \cline{2-3}
2.00-2.50 & 1.58 &75 & 1.71 & 41 & 1.61 & 57 & 1.53 &  9 & 1.63 &  9 & & & & & & \\\cline{8-9}
2.50-3.00 & 2.08 &26 & 2.08 & 21 &      &    & & & 1.99 & 4 & & & & & & \\ 
3.00-3.50 & 2.50 &11 & 2.24 &  2 & & & & & & & & & & & & \\
\hline
Total & & 159 & & 255 & & 373 & & 393 & & 405 & & 638 & & 547 & &  120\\
\hline
\end{tabular}  
}
\end{table*}

\begin{table}
\center
\caption{Number of stars for different absolute magnitude intervals for thick 
disk (symbols as in Table 4).}
{\scriptsize
\begin{tabular}{rrrrrr}
\hline
$M(g^{'})\rightarrow$& (4-5] & (5-6]& (6-7]&(7-8] & (8-9]\\
\hline
$r_{r}-r_{2}$ & $z^{*}$~~~N & $z^{*}$~~~N & $z^{*}$~~~N & $z^{*}$~~~N & $z^{*}$~~~N \\
\hline
0.5-1.0 &          &          &           &         & 0.64~~30 \\
1.0-1.5 & 1.05~~17 &          & 1.13~~~1 & 1.04~~10 & 0.95~~59 \\ \cline{2-2} \cline{4-4} \cline{5-5}
1.5-2.0 & 1.29~~51 &          & 1.30~~95 & 1.38~~40 & 1.37~~79 \\ \cline{3-3}
2.0-2.5 & 1.66~~69 & 1.78~~80 & 1.66~136 & 1.74~~65 & 1.69~~84 \\ \cline{6-6}
2.5-3.0 & 2.05~~44 & 2.05~116 & 2.06~124 & 2.05~~99 & 2.02~~44 \\ 
3.0-3.5 & 2.42~~20 & 2.42~115 & 2.43~129 & 2.42~~67 & 2.42~~23 \\
3.5-4.0 & 2.77~~20 & 2.81~107 & 2.80~104 & 2.82~~53 & 2.78~~10 \\
4.0-4.5 & 3.14~~18 & 3.18~~63 & 3.17~~72 & 3.16~~47 & 3.09~~~7 \\
4.5-5.0 & 3.48~~~3 & 3.57~~66 & 3.54~~69 & 3.50~~30 & 3.57~~~4 \\ \cline{5-5}
5.0-5.5 &          & 3.91~~51 & 3.90~~54 & 3.75~~~5 &          \\ 
5.5-6.0 &          & 4.30~~38 & 4.26~~53 &          &          \\
6.0-6.5 &          & 4.66~~34 & 4.62~~44 &          &          \\
6.5-7.0 &          & 5.06~~11 & 5.03~~38 &          &          \\ \cline{4-4}
7.0-8.0 &          & 5.27~~~4 & 5.38~~58 &          &          \\ 
8.0-9.0 &          &          & 5.83~~24 &          &          \\
\hline
Total   &      242 &       685 &      1001 &      416 &      340 \\
\hline
\end{tabular}
}  
\end{table}

\begin{table}
\center
\caption{Number of stars for different absolute magnitude intervals for halo 
(symbols as in Table 4).}
{\scriptsize
\begin{tabular}{rrrrrrrrr}
\hline
$M(g^{'})\rightarrow$ & \multicolumn{2} {c} {(4-5]} & \multicolumn{2} {c} {(5-6]} 
&\multicolumn{2} {c} {(6-7]} & \multicolumn{2} {c} {(7-8]}\\
\hline
$r_{1}-r_{2}$ & $z^{*}$ &    N &$z^{*}$&  N &$z^{*}$ & N & $z^{*}$ & N \\
\hline
2-4  &  2.58 & 19 &      &     &      &     & 2.73 & 14 \\ \cline{2-3} \cline{8-9}
4-6  &  3.81 & 34 &      &     &      &     & 4.17 & 42 \\
6-8  &  5.05 & 25 & 5.37 &  99 & 5.38 &  76 & 5.04 & 48 \\
8-10 &  6.57 & 10 & 6.71 & 113 & 6.65 & 117 & 6.72 & 10 \\
10-15&  9.10 & 27 & 9.13 & 178 & 8.73 & 129 &      &    \\ \cline{6-7}
15-20& 12.97 & 17 &12.54 &  57 &      &     &      &    \\ \cline{4-5}
20-25& 17.00 &  3 &16.02 &   2 &      &     &      &    \\ 
25-30& 19.44 &  4 &      &     &      &     &      &    \\ \cline{2-3}
30-35& 22.11 &  2 &      &     &      &     &      &    \\
\hline
Total&       &141 &      & 449 &      & 322 &      &114 \\
\hline
\end{tabular}
}  
\end{table}

\begin{figure}
\resizebox{7cm}{4cm}{\includegraphics*{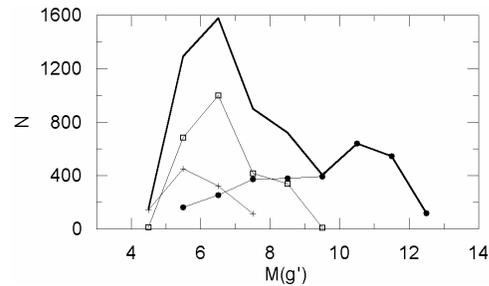}} 
\caption[] {Absolute magnitude ranges dominated by different populations. 
Symbols: (+) halo, ($\sq$) thick disk, ($\bullet$) thin disk, and thick line: 
the distribution of stars for all absolute magnitudes}
\end{figure}

\begin{eqnarray}
(g^{'}-M(g^{'}))_{o} =  5 \log r_{l} - 5\\
z_{l}=r_{l} sin b					
\end{eqnarray}
where $g_{0}^{'}$ is the limiting apparent magnitude (17 and 22, for the bright 
and faint stars, respectively), $r_{l}$ the limiting distance of completeness relative 
to the Sun, and $M(g_{0}^{'})$ the appropriate absolute magnitude $M_{1}$ or $M_{2}$ 
for the absolute magnitude interval $M_{1}-M_{2}$ considered.

\begin{table*}
\center
\caption{Galactic model parameters for different absolute magnitude 
intervals for thin disk resulting from the comparison of observed 
logarithmic space densities with a (unique) density law (Fig. 9). The 
columns give: $M(g^{'})$ absolute magnitude interval, density law, 
logarithmic local space density $n^{*}$, scale height for $sech$ or $sech^{2}$ 
density law $z_{0}$, scale height for exponential density law H, 
$\chi^{2}$, standart deviation $s$, and local space densities for 
Hipparcos $\odot$.}  
{\scriptsize
\begin{tabular}{cccccrc}
\hline
$M(g^{'})$ & Density Law & $n^{*}$ &  $z_{o}/ H$ (pc)&  $\chi^{2}.(10^{-10})$&$s$& $\odot$ \\
\hline
(12-13] &      $exp$ &$8.40^{+0.07}_{-0.07}$ &$101^{+11}_{-7}$ &      4362816 & $\pm$ 0.14 & 8.05\\
        &     $sech$ &$8.14^{+0.07}_{-0.07}$ &$98/162^{+17}_{-13}$ &  3767270 & 0.14 & \\
        & $sech^{2}$ &$8.08^{+0.06}_{-0.06}$ &$166/275^{+22}_{-22}$ & 2691626 & 0.12 & \\
(11-12] &      $exp$ &$8.98^{+0.02}_{-0.05}$ &$103^{+3}_{-4}$ &      1940921 &  0.10 & 7.92\\
        &     $sech$ &$8.69^{+0.04}_{-0.04}$ &$102/169^{+7}_{-5}$ &  1690410 &  0.09 & \\
        & $sech^{2}$ &$8.55^{+0.02}_{-0.03}$ &$188/312^{+6}_{-9}$ &   755904 &  0.06 & \\
(10-11] &      $exp$ &$8.38^{+0.04}_{-0.04}$ &$168^{+7}_{-6}$&        628271 &  0.10 & 7.78\\
        &     $sech$ &$8.11^{+0.03}_{-0.06}$ &$168/278^{+9}_{-16}$ &  857229 &  0.10 & \\
        & $sech^{2}$ &$7.99^{+0.04}_{-0.04}$ &$300/497^{+23}_{-21}$ & 994846 &  0.06 & \\
 (9-10] &      $exp$ &$7.60^{+0.05}_{-0.05}$ &$264^{+14}_{-13}$&      274519 &  0.10 & 7.63\\
        &     $sech$ &$7.34^{+0.05}_{-0.05}$ &$256/424^{+24}_{-21}$ & 295956 &  0.06 & \\
        & $sech^{2}$ &$7.23^{+0.06}_{-0.06}$ &$460/762^{+50}_{-41}$ & 370010 &  0.06 & \\
  (8-9] &      $exp$ &$7.47^{+0.07}_{-0.06}$ &$292^{+22}_{-13}$ & 247010 &      0.13 & 7.52\\
  (7-8] &      $exp$ &$7.50^{+0.02}_{-0.02}$ &$312^{+3}_{-4}$&      8207 &      0.02 & 7.48\\
  (6-7] &      $exp$ &$7.43^{+0.02}_{-0.02}$ &$326^{+5}_{-5}$&      6295 &      0.03 & 7.47\\
  (5-6] &      $exp$ &$7.43^{+0.01}_{-0.01}$ &$334^{+2}_{-2}$&       162 &      0.01 & 7.47\\
\hline
\end{tabular}
}  
\end{table*}

\section{Galactic model parameters}

\subsection{Density laws for different populations and different absolute 
magnitudes: new approach for the model parameter estimation} 

In the literature different density laws, such as secans hyperbolicus ($sech$), 
squared secans hyperbolicus ($sech^{2}$) or exponentials, were used for 
parameterization of thin disk data, whereas the exponential density law was 
sufficient for thick disk data. In the present study, we compared the observed logarithmic 
space densities for thin disk with all density laws cited above. It turned out 
that $sech^{2}$ law fits better for three faint absolute magnitude intervals, 
i.e. $12<M(g^{'})\leq13$, $11<M(g^{'})\leq12$, and $10<M(g^{'})\leq11$, whereas 
exponential law is favourable for brighter absolute magnitudes (Table 7 and 
Fig. 9). The argument used for this conclusion is the difference between the local 
space density resulting from the comparison of the observed space density functions 
with the density laws and the Hipparcos one (Jahreiss \& Wielen\ 1997). Table 7 
gives also the corresponding scale-heights. It is interesting, the scale-height 
increase monotonously when one goes from the faint magnitudes towards the bright 
ones, however there is a discontunity at the transition region of two density laws. 
The range and the mean of the scale-height for thin disk are 264-497 pc, and 
327 pc, respectively. Although 497 pc seems an extreme value, it is close to the 
upper limit cited by Robin \& Cr\'{e}z\'{e}\ (1986). 

\begin{figure*}
\resizebox{16cm}{9cm}{\includegraphics*{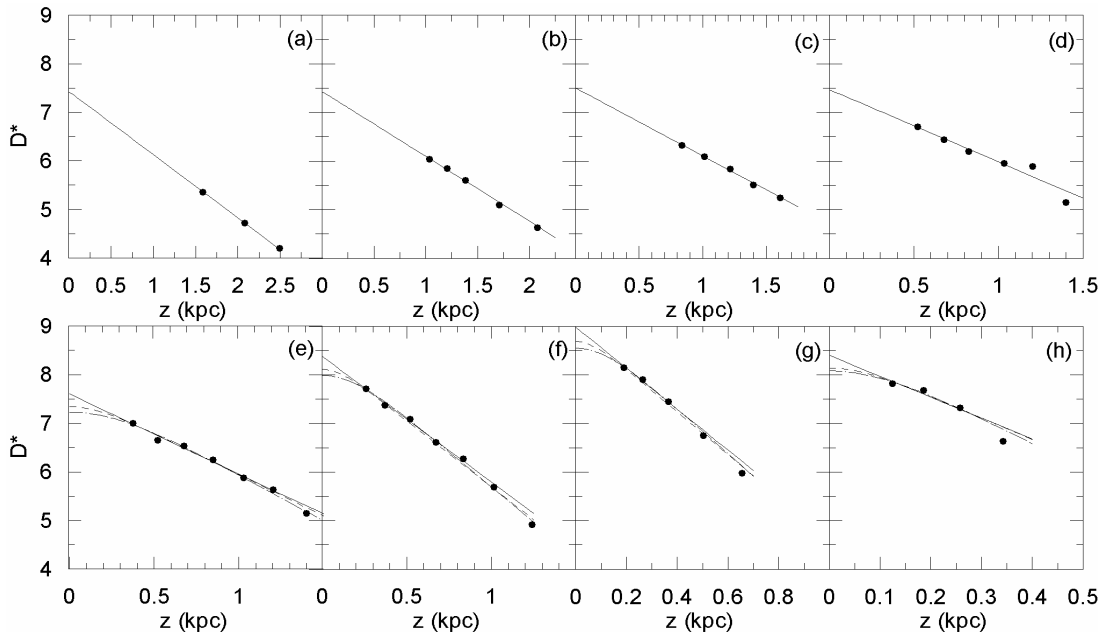}} 
\caption[] {Comparison of the observed space density function with the density 
laws for different absolute magnitude intervals for the thin disk. (a) (5-6], 
(b) (6-7], (c) (7-8], (d) (8-9], (e) (9-10], (f) (10-11], (g) (11-12], and
(h) (12-13]. Continuous curve: represents exponential law, whereas dashed 
curve, $sech$ law and dashed-dot curve, $sech^{2}$ law.}
\end{figure*}

\begin{table}
\center
{\scriptsize
\caption{Galactic model parameters for thick disk. $n_{2}/n_{1}$ indicates the 
local space density for thick disk relative to thin disk. Other symbols are 
same as in Table 7.}
\begin{tabular}{ccccrc}
\hline
$M(g^{'})$ & $n^{*}$ & $z_{o}/H$ (pc) & $\chi^{2}.(10^{-10})$ & $s$ & $n_{2}/n_{1}(\%)$\\
\hline
(8-9] &$6.19^{+0.01}_{-0.01}$ &$970^{+27}_{-29}$ & 2002 & $\pm$ 0.02 & 5.25 \\
(7-8] &$6.31^{+0.07}_{-0.08}$ &$806^{+52}_{-47}$ & 18505&       0.07 & 6.46 \\
(6-7] &$6.42^{+0.03}_{-0.03}$ &$895^{+18}_{-23}$ & 5879 &       0.06 & 9.77 \\
(5-6] &$6.41^{+0.04}_{-0.04}$ &$876^{+26}_{-25}$ & 5503 &       0.08 & 9.55 \\
\hline
\end{tabular}
}  
\end{table}

\begin{table}
\center
\caption{Galactic model parameters for halo. $\kappa$ and $n_{3}/n_{1}$ 
give the axial ratio and the local space density for halo relative to thin disk, 
respectively. Other symbols are as in Table 7.}
{\scriptsize
\begin{tabular}{ccccrc}
\hline
$M(g^{'})$& $n^{*}$ & $\kappa$ &$\chi^{2}.(10^{-10})$&  $s$ & $n_{3}/n_{1}(\%)$ \\
\hline
(7-8] &$3.98^{+0.15}_{-0.14}$ & $0.78^{+0.22}_{-0.18}$ & 1959 & $\pm$ 0.30 & 0.02 \\
(6-7] &$4.27^{+0.10}_{-0.11}$ & $0.71^{+0.10}_{-0.08}$ &  778 & 0.13 & 0.07 \\
(5-6] &$4.53^{+0.06}_{-0.06}$ & $0.57^{+0.03}_{-0.03}$ &  271 & 0.14 & 0.13 \\
(4-5] &$4.79^{+0.07}_{-0.06}$ & $0.26^{+0.01}_{-0.01}$ &  189 & 0.16 & 0.31 \\
\hline
\end{tabular}
}  
\end{table}

For the thick disk, the observed logarithmic space density functions are compared 
with the exponential density law for the absolute magnitude intervals 
$8<M(g^{'})\leq9$, $7<M(g^{'})\leq8$, $6<M(g^{'})\leq7$, and $5<M(g^{'})\leq6$ 
(Table 8, Fig. 10). The scale-height for different absolute magnitude intervals 
varies from 806 pc to 970 pc and are in agreement with the recent values in the cited 
literature (see the references in Table 1). The local space density relative to the 
one of thin disk, for the corresponding absolute magnitude intervals in Table 7, 
increases from the faint absolute magnitudes to the bright ones in a range 
(5.25-9.77)\%, again in agreeable with the literature. The observed logarithmic 
space density functions for halo have been compared with the de Vaucouleurs 
density law for the absolute magnitude intervals $7<M(g^{'})\leq8$, 
$6<M(g^{'})\leq7$, $5<M(g^{'})\leq6$, and $4<M(g^{'})\leq5$ (Table 9 and Fig. 
11). The axial ratio decreases from relative absolute magnitudes to the bright 
ones, whereas the local space density relative to the thin disk for the 
corresponding absolute magnitude intervals increases in the same order. The 
parameters cited here are in the range given in the literature, except the 
ones for the interval $4<M(g^{'})\leq5$. The parameters derived for three 
populations have been tested by the luminosity function, given in Table 10 
and Fig. 12, where $\varphi^{*}(M)$ is the total of the local space densities 
for three populations. There is a good agreement between our luminosity function 
and that of Hipparcos (Jahreiss \& Wielen\ 1997) with the exception of absolute 
magnitude interval $11<M(g^{'})\leq12$. Also, the corresponding standard 
deviations for all intervals are small (Table 10). 

\begin{table}
\center
\caption{Luminosity function resulting from the combination of local space 
densities for three populations, i.e. thin and thick disks, and halo, taken 
from Tables 7-9. The luminosity function of Hipparcos is given in the last 
column.}
\begin{tabular}{ccrc}
\hline
$M(g^{'})$ & $\varphi^{*}(M)$ & $s$ & $\odot$\\
\hline
(12-13]& 8.08 & $\pm$0.14 & 8.05\\
(11-12]& 8.55 & 0.10 & 7.92\\
(10-11]& 7.99 & 0.10 & 7.78\\
(9-10] & 7.60 & 0.10 & 7.63\\
(8-9]  & 7.49 & 0.13 & 7.52\\
(7-8]  & 7.53 & 0.02 & 7.48\\
(6-7]  & 7.47 & 0.03 & 7.47\\
(5-6]  & 7.47 & 0.01 & 7.47\\
\hline
\end{tabular}  
\end{table}

\begin{figure}
\resizebox{8cm}{8.8cm}{\includegraphics*{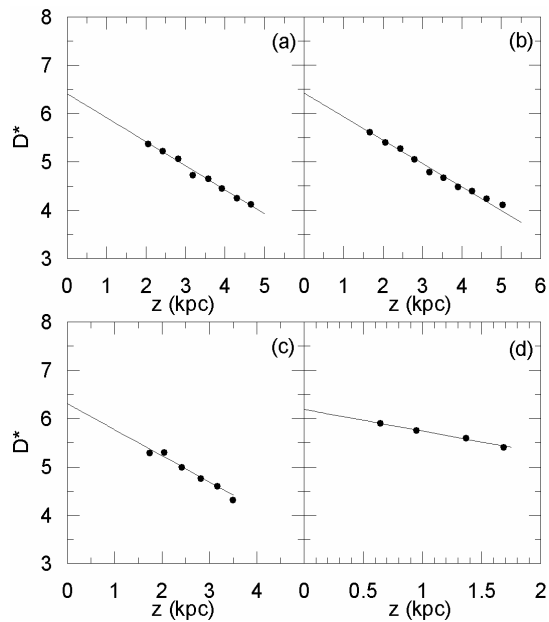}} 
\caption[] {Comparison of the observed space density function with the 
exponential density law for different absolute magnitude intervals for 
thick disk. (a) (5-6], (b) (6-7], (c) (7-8], and (d) (8-9].}
\end{figure}

\begin{figure}
\resizebox{8cm}{8.8cm}{\includegraphics*{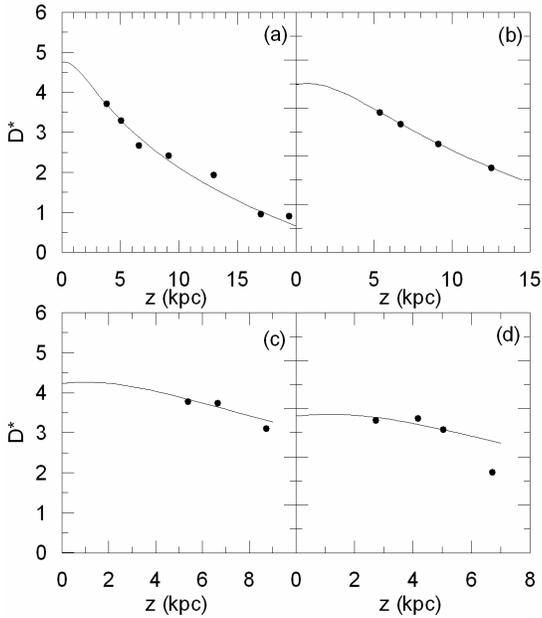}} 
\caption[] {Comparison of the observed space density function with the de 
Vaucouleurs density law for different absolute magnitude intervals for halo.
(a) (4-5], (b) (5-6], (c) (6-7], and (d) (7-8].}
\end{figure}

\begin{figure}
\resizebox{8cm}{5cm}{\includegraphics*{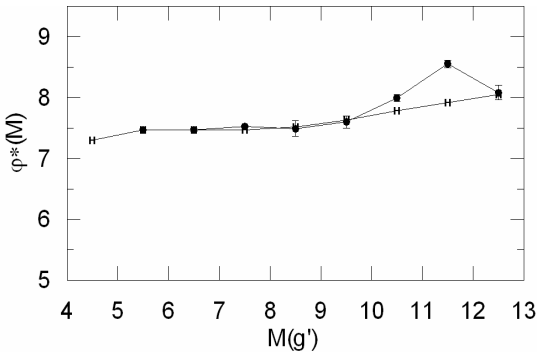}} 
\caption[] {Luminosity function obtained from combination of the local space 
densities for thin and thick disks and  halo, resulting from comparison of the 
observed space density function with the density laws, for different absolute 
magnitude intervals. (H) shows the Hipparcos values.}
\end{figure}

We used the procedure of Phleps et al.\ (2000) for the error estimation in 
Tables 7-9 (above) and Tables 12 and 14 (in the following sections), i.e. 
by changing the values of the parameters until $\chi^{2}$ increases or 
decreases by 1.   

\subsection{Model parameter estimation by the procedure in situ}

We estimated the local space density and scale-height for thin disk and 
thick disk, and the local space density and axial ratio for halo 
simultaneously by the procedure in situ, i.e. by comparison of the combined 
observed space density functions with the combined density laws. We carried 
out this work for two sets of absolute magnitude intervals, $5<M(g^{'})\leq10$ 
and $5<M(g^{'})\leq13$. The second set covers the thin disk stars with 
$10<M(g^{'})\leq13$ whose density functions behave differently than the density 
functions for stars with other absolute magnitudes. Number of stars as a 
function of distance $r$ relative to the Sun for eight absolute magnitude 
intervals are given in Table 11 and the density functions per unit absolute 
magnitude interval evaluated by these data are shown in Table 12 and Table 13 
for the sets $5<M(g^{'})\leq10$ and $5<M(g^{'})\leq13$, respectively.     

\subsubsection {Model parameters by means of absolute magnitudes 
$5<M(g^{'})\leq10$}

The observed space densities per absolute magnitude interval for three 
populations, i.e. thin and thick disks, and halo for stars with 
$5<M(g^{'})\leq10$, (Table 12) are compared with the combined density 
laws in situ (Fig. 13). The derived  parameters are given in Table 14. All 
these parameters are in agreement with the ones given in Table 1. Especially 
the relative local space density for thick disk, 8.32\%, lies within the range 
given in two recent works (Chen et al.\ 2001 and Siegel et al.\ 2002). The 
resulting luminosity function (Fig. 14) from the comparison of the model with 
these parameters and the combined observed density functions per absolute 
magnitude interval is also in agreement with the one of Hipparcos (Jahreiss 
\& Wielen\ 1997). However the error bars are longer than the ones in Fig. 12, 
particularly for the faint magnitudes.    

\begin{figure}
\resizebox{8cm}{5cm}{\includegraphics*{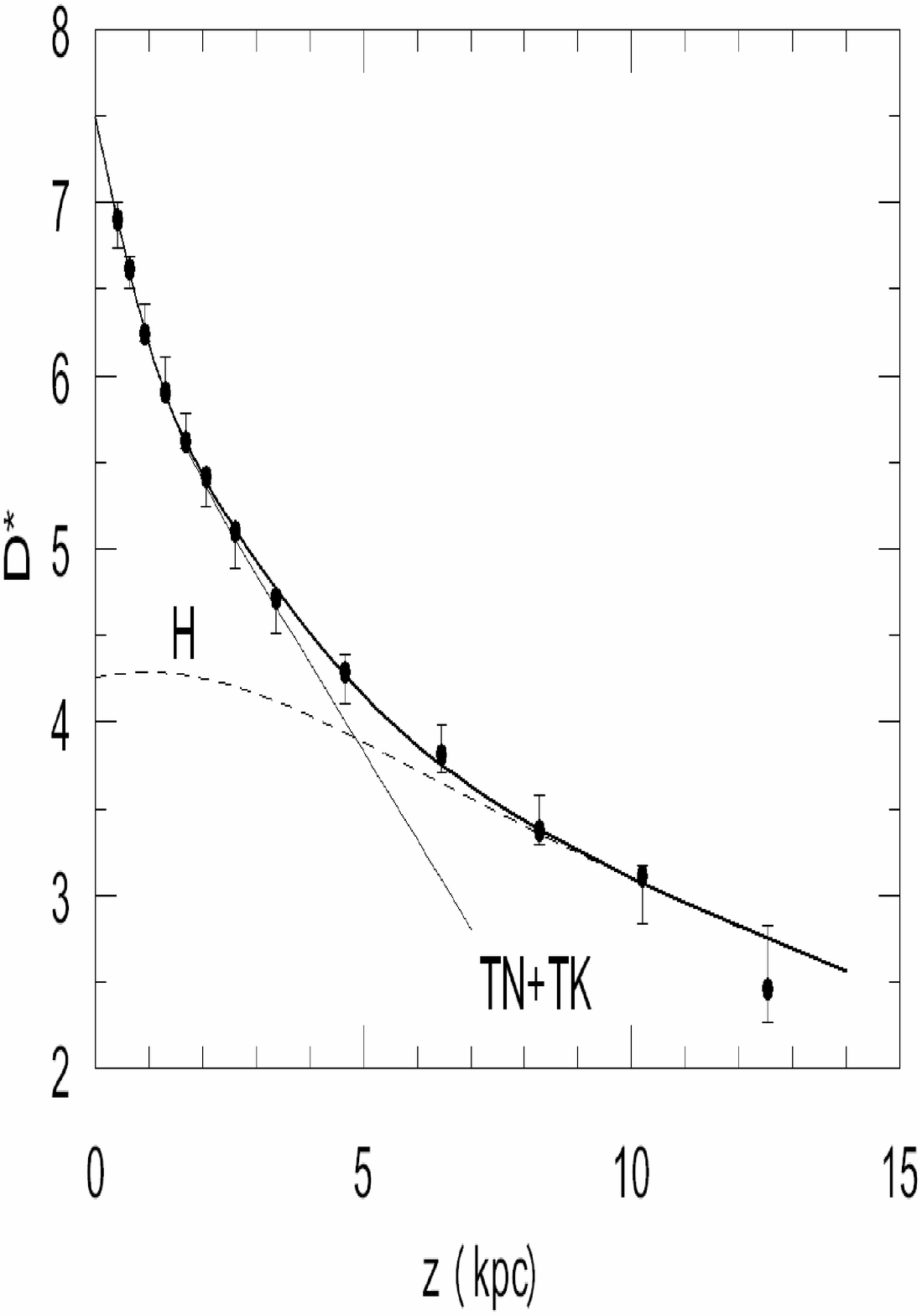}} 
\caption[] {Comparison of the observed space density function combined for 
thin and thick disks and halo with the combined density law, for stars with 
$5<M(g^{'})\leq10$.}
\end{figure}

\begin{figure}
\resizebox{8cm}{5cm}{\includegraphics*{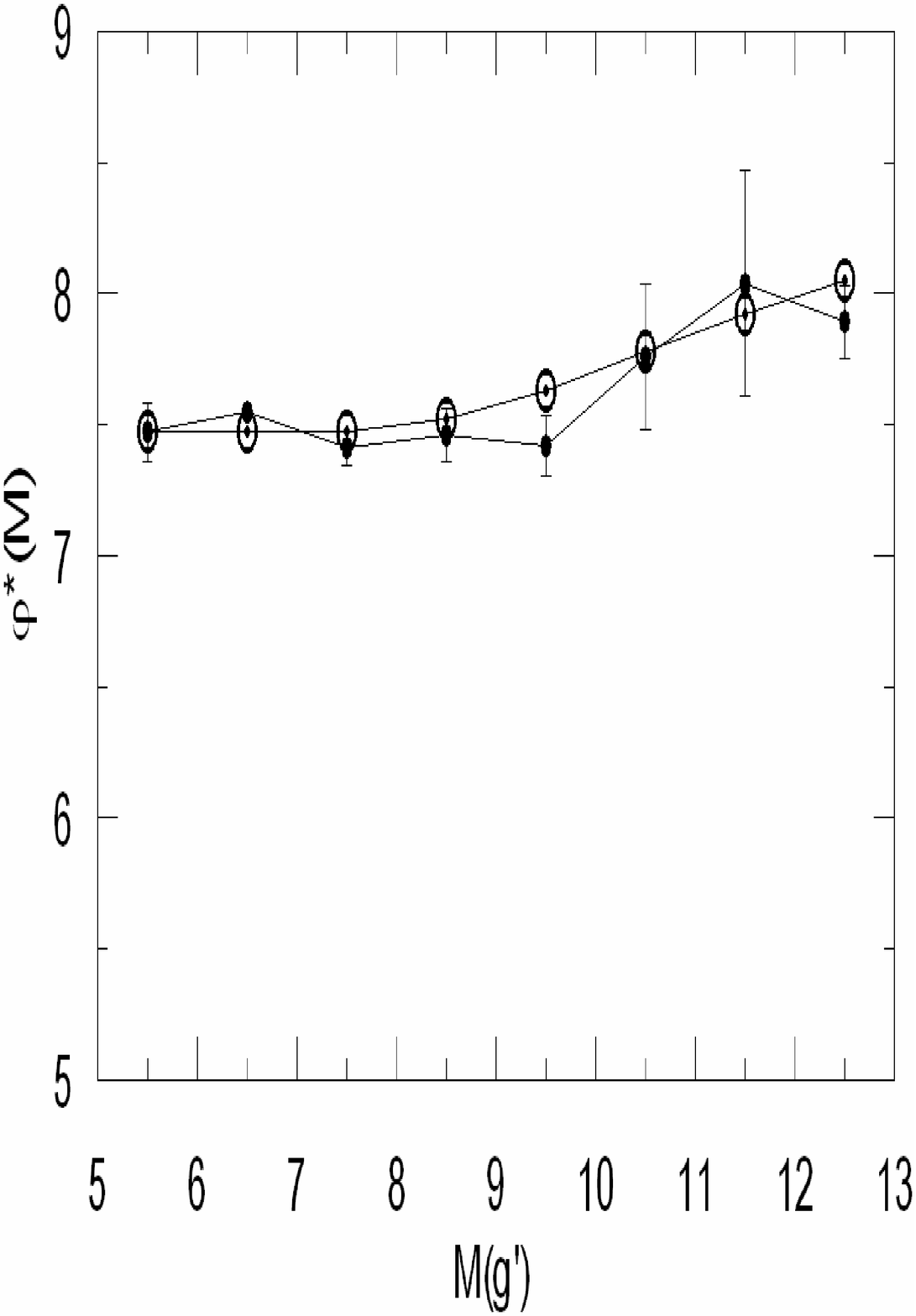}} 
\caption[] {Luminosity function resulting from the comparison of the combined 
observed space density function with the combined density law, for stars with 
$5<M(g^{'})\leq10$.}
\end{figure}

\begin{table*}
\center
\caption{Number of stars as a function of distance $r$ relative to the Sun, and 
the corresponding mean distance $z^{*}$ from the Galactic plane, for different 
absolute magnitude intervals (distances in kpc). Horizontal thick lines 
correspond the limiting distance of completeness.}
{\scriptsize
\begin{tabular}{rrrrrrrrrrrrrrrrr}
\hline
$M(g^{'})\rightarrow$  & \multicolumn{2} {c} {(5-6]} & \multicolumn{2} {c} {(6-7]}& \multicolumn{2} {c} {(7-8]} &\multicolumn{2} {c} {(8-9]} &\multicolumn{2} {c} {(9-10]} & \multicolumn{2} {c} {(10-11]}& \multicolumn{2} {c} {(11-12]} & \multicolumn{2} {c} {(12-13]} \\
\hline
$r_{1}-r_{2}$ & $z^{*}$ & N & $z^{*}$ & N & $z^{*}$ & N & $z^{*}$ & N & $z^{*}$ & N & $z^{*}$ & N & $z^{*}$ & N &    $z^{*}$ & N \\
\hline
0.0-0.2 &  & & & & & & & & & 0.15 & 1 & 0.13 & 30 & 0.12 & 20 \\ \cline{14-17} 
0.2-0.4 &  & & & & & & & & 0.26 & 25 & 0.24 & 125 & 0.23 & 244 & 0.22 & 72 \\ \cline{10-13} \cline{16-17} 
0.4-0.7 &  & & & & & & 0.45 & 77 & 0.41 & 96 & 0.41 & 236 & 0.39 & 230 & 0.34 & 28  \\ \cline{8-9} \cline{14-15}
0.7-1.0 &  & & & & 0.65 & 57 & 0.63 & 135 & 0.64 & 100 & 0.62 & 163 & 0.59 & 41 & & \\ \cline{6-7} \cline{12-13}
1.0-1.5 &  & & 1.00 & 86 & 0.93 & 173 & 0.93 & 179 & 0.92 & 118 &  0.89 & 109 & 0.75 & 2 & &\\ \cline{4-5} \cline{10-11}
1.5-2.0 & 1.39 & 47 & 1.29 & 201 & 1.32 & 136 & 1.30 & 149 & 1.26 & 53 & 1.27 & 4 & & & &  \\ \cline{2-3}
2.0-2.5 & 1.69 & 158& 1.68 & 177 & 1.68 & 122 & 1.67 &  93 & 1.63 &  9 &      &   & & & &  \\ \cline{8-9}
2.5-3.0 & 2.06 & 142& 2.06 & 145 & 2.05 &  99 & 2.02 &  44 & 1.99 &  4 &      &   & & & &  \\
3.0-4.0 & 2.60 & 233& 2.59 & 235 & 2.61 & 134 & 2.53 &  33 &      &    &      &   & & & &  \\ \cline{6-7}
4.0-5.0 & 3.38 & 129& 3.35 & 141 & 3.29 &  77 & 3.21 &  10 &      &    &      &   & & & &  \\ \cline{4-5}
5.0-7.5 & 4.63 & 209& 4.68 & 284 & 4.54 &  91 & 3.81 &   1 &      &    &      &   & & & &  \\
7.5-10.0& 6.53 & 141& 6.40 & 181 & 6.42 &  14 &      &     &      &    &      &   & & & &  \\
10.0-12.5&8.28 & 100& 8.27 &  93 &      &     &      &     &      &    &      &   & & & &  \\
12.5-15.0&10.21&  78& 9.90 &  36 &      &     &      &     &      &    &      &   & & & &  \\
15.0-20.0&12.54&  57&      &     &      &     &      &     &      &    &      &   & & & &  \\\cline{2-3}
20.0-25.0&16.02&   2&      &     &      &     &      &     &      &    &      &   & & & &  \\
\hline
Total & & 1296 & & 1579 & & 903 & & 721 & & 405 & & 638 & & 547 &  & 120 \\
\hline
\end{tabular}  
}
\end {table*}

\subsubsection {Model Parameters by means of absolute magnitudes 
$5<M(g^{'})\leq13$}

We carried out the work cited in the previous paragraph for stars with a 
larger range of absolute magnitude, i.e. $5<M(g^{'})\leq13$. The  
observed density function is given in Table 13 and its comparison with the 
combined density law is shown in Fig. 15. The derived parameters (Table 15), 
especially the local densities, are rather different than the ones cited in 
Sections (6.1) and (6.2.1). The reason for this discrepancy is that stars 
with absolute magnitudes $10<M(g^{'})\leq13$ have relatively larger local 
space densities (Hipparcos; Jahreiss \& Wielen\ 1997) and are closer to the 
Sun relative to stars brighter than $M(g^{'})$=10, and they affect the 
combined density function considerably. Also the corresponding luminosity 
function is not in agreement with the one of Hipparcos, except one absolute 
magnitude interval, $12<M(g^{'})\leq13$ (Fig. 16).

\begin{table}
\center
\caption{Logarithmic space density function, $D^{*}=\log D+10$, per unit 
absolute magnitude interval for stars with $5<M(g^{'})\leq10$. $<N>$ is 
the weighted mean of number of stars for the absolute magnitude intervals 
in Table 11 under consideration, the other symbols are explained in the text 
(distances in kpc, volumes in pc$^{3}$).}
\begin{tabular}{ccccc}
\hline
$r_{1}-r_{2}$ & $\Delta V_{1,2}$ &    $z^{*}$ & $<N>$ &    $D^{*}$ \\
\hline
   0.4-0.7 &   1.20 (5) &      0.408 &         96 &       6.90 \\
   0.7-1.0 &   2.83 (5) &      0.634 &        118 &       6.62 \\
   1.0-1.5 &   1.02 (6) &      0.929 &        176 &       6.24 \\
   1.5-2.0 &   1.99 (6) &      1.302 &        162 &       5.91 \\
   2.0-2.5 &   3.28 (6) &      1.678 &        138 &       5.62 \\
   2.5-3.0 &   4.90 (6) &      2.055 &        129 &       5.42 \\
   3.0-4.0 &   1.59 (7) &      2.600 &        201 &       5.10 \\
   4.0-5.0 &   2.63 (7) &      3.365 &        135 &       4.71 \\
   5.0-7.5 &   1.28 (8) &      4.656 &        247 &       4.29 \\
  7.5-10.0 &   2.49 (8) &      6.457 &        161 &       3.81 \\
 10.0-12.5 &   4.10 (8) &      8.279 &         97 &       3.37 \\
 12.5-15.0 &   6.12 (8) &     10.208 &         78 &       3.11 \\
 15.0-20.0 &   1.99 (9) &     12.536 &         57 &       2.46 \\
\hline
\end{tabular}  
\end{table}

\begin{table}
\center
\caption{Logarithmic space density function, $D^{*}=\log D+10$, per unit 
absolute magnitude interval for stars with $5<M(g^{'})\leq13$. $<N>$ is 
the weighted mean of number of stars for the absolute magnitude intervals 
in Table 11 under consideration, the other symbols are explained in the text 
(distances in kpc, volumes in pc$^{3}$).}
\begin{tabular}{rrrrr}
\hline
$r_{1}-r_{2}$ & $\Delta V_{1,2}$ &    $z^{*}$ & $<N>$ &    $D^{*}$ \\
\hline
     0-0.2 &   3.44 (3) &      0.127 &         25 &       7.86 \\
   0.2-0.4 &   2.41 (4) &      0.230 &        147 &       7.79 \\
   0.4-0.7 &   1.20 (5) &      0.402 &        187 &       7.19 \\
   0.7-1.0 &   2.83 (5) &      0.628 &        133 &       6.67 \\
   1.0-1.5 &   1.02 (6) &      0.929 &        176 &       6.24 \\
   1.5-2.0 &   1.99 (6) &      1.302 &        162 &       5.91 \\
   2.0-2.5 &   3.28 (6) &      1.678 &        138 &       5.62 \\
   2.5-3.0 &   4.90 (6) &      2.055 &        129 &       5.42 \\
   3.0-4.0 &   1.59 (7) &      2.600 &        201 &       5.10 \\
   4.0-5.0 &   2.63 (7) &      3.365 &        135 &       4.71 \\
   5.0-7.5 &   1.28 (8) &      4.656 &        247 &       4.29 \\
  7.5-10.0 &   2.49 (8) &      6.457 &        161 &       3.81 \\
 10.0-12.5 &   4.10 (8) &      8.279 &         97 &       3.37 \\
 12.5-15.0 &   6.12 (8) &     10.208 &         78 &       3.11 \\
 15.0-20.0 &   1.99 (9) &     12.536 &         57 &       2.46 \\
\hline
\end{tabular}  
\end{table}

\begin{table}
\center
\caption{Galactic model parameters estimated by comparison of the 
logarithmic space density function  for stars with $5<M(g^{'})\leq10$ (given 
in Table 12) and the combined density laws (Fig.13). The symbols give: 
$n^{*}$: logaritmic local space density, H: scale height, $n/n_{1}$: 
local space density relative thin disk, and $\kappa$= c/a: axial ratio 
for halo.}
\begin{tabular}{cccc}
\hline
Parameter & Thin disk & Thick disk & Halo\\
\hline
$n^{*}$   & $7.46^{+0.02}_{-0.02}$ & $6.38^{+0.03}_{-0.03}$  & $4.30^{+0.29}_{-0.45}$ \\
$H$ (pc)  & $275^{+6}_{-5}$ & $851^{+43}_{-41}$ & $--$ \\
$n/n_{1}($\%$)$ & $--$ & $8.32$ & $0.07$ \\
$\kappa$  & $--$ & $--$& $0.67^{+0.33}_{-0.23}$\\
\hline
\end{tabular}  
\end{table}

\begin{table}
\center
\caption{Galactic model parameters estimated by comparison of the 
logarithmic space density function for stars with $5<M(g^{'})\leq13$ 
(given in Table 13) and the combined density laws (Fig. 14). Symbols 
as in Table 14.}
\begin{tabular}{cccc}
\hline
Parameter & Thin disk & Thick disk & Halo\\
\hline
$n^{*}$   & 8.21 & 6.22 & 4.30 \\
$H$ (pc)  & 184  & 1048 & $--$ \\
$n/n_{1}($\%$)$  & $--$ & 1.00 & $0.01$ \\
$\kappa$  & $--$ & $--$ & 0.67\\
\hline
\end{tabular}  
\end{table}

\begin{figure}
\resizebox{8cm}{5cm}{\includegraphics*{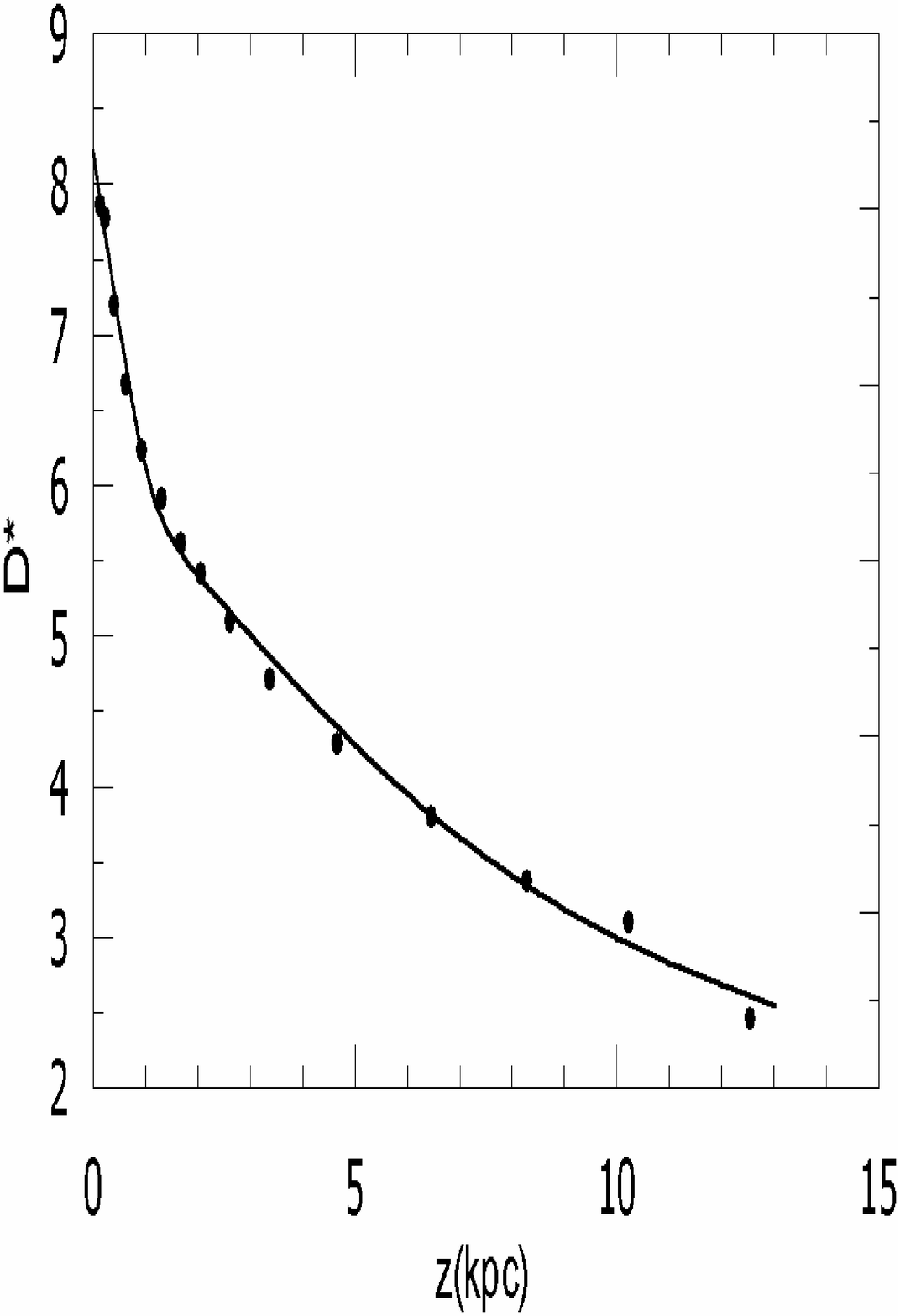}} 
\caption[] {Comparison of the observed space density function combined for 
thin and thick disks and halo with the combined density laws, for stars with 
$5<M(g^{'})\leq13$.}
\end{figure}

\begin{figure}
\resizebox{8cm}{5cm}{\includegraphics*{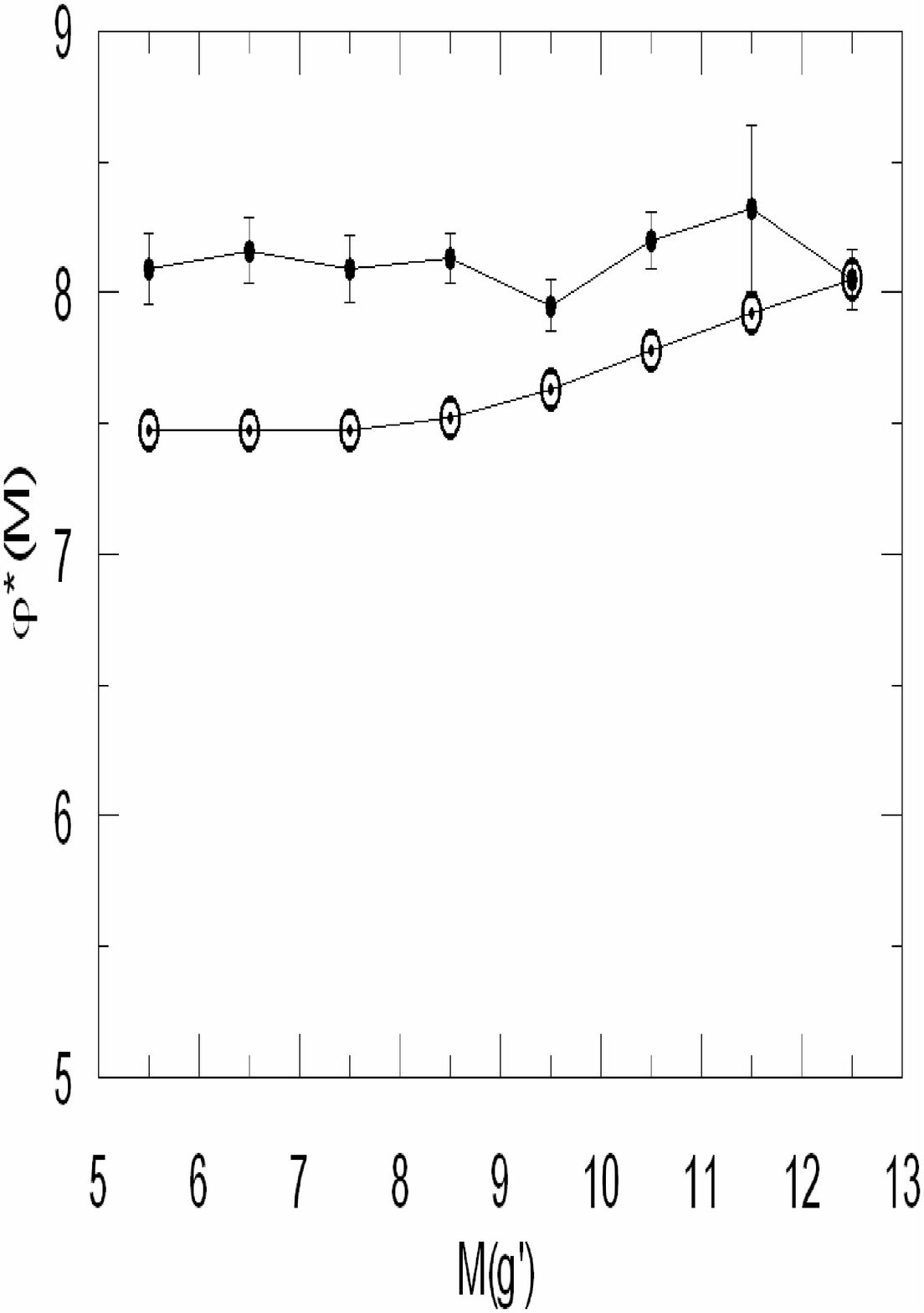}} 
\caption[] {Luminosity function resulting from the comparison of the combined 
observed space density function with the combined density law, for stars with  
$5<M(g^{'})\leq13$.}
\end{figure}

\section {Discussion}

We estimated the Galactic model parameters by means of the vertical density 
distribution for the field SA 114 ($\alpha=22^{h}40^{m}00^{s},\delta= 
00^{o}00^{'}00{''}$; $l=68^{o}.15$, $b=-48^{o}.38$; 4.239 square-degree; epoch 
2000) by means of  two procedures, i.e. a new approach and the procedure in situ. The 
new approach is based on the comparison of the observed space density functions 
per absolute magnitude interval with a unique density law for each population 
individually, where the separation of stars into different population types is 
carried out by a slight modification of the method of Karaali\ (1994), i.e. by 
their spatial distribution as a function of absolute and apparent magnitude. 
This approach covers nine absolute magnitude intervals, i.e. $4<M(g^{'})\leq5$, 
$5<M(g^{'})\leq6$, $6<M(g^{'})\leq7$, $7<M(g^{'})\leq8$, $8<M(g^{'})\leq9$, 
$9<M(g^{'})\leq10$, $10<M(g^{'})\leq11$, $11<M(g^{'})\leq12$, and 
$12<M(g^{'})\leq13$. Whereas the procedure in situ compares the observed space 
density functions per absolute magnitude interval with the combination of a 
set of density law representing all the populations. This procedure is carried 
out for two absolute magnitude intervals, $5<M(g^{'})\leq10$ and 
$5<M(g^{'})\leq13$. However, we will not discuss the parameters for 
$5<M(g^{'})\leq13$, since they are rather different than the parameters 
appeared in the literature up to now. 

\subsection{Parameters determined by means of the new approach}

The new approach provides absolute magnitude dependent Galactic model 
parameters. The scale-height for thin disk increases monotonously from 
the faint magnitudes to bright ones, however there is a 
discontunity at the transition region of two density laws (exponential law 
for $M(g^{'})\leq10$, and $sech^{2}$ law for $10<M(g^{'})\leq13$). The range 
of this parameter is 264-497 pc, and one can find all the values derived 
for different absolute magnitude intervals in the literature including 
the extreme one, i.e. 497 pc which is close to the one cited by Robin \& 
Cr\'{e}z\'{e}\ (1986). The thick disk is dominant in the intervals 
$5<M(g^{'})\leq6$, $6<M(g^{'})\leq7$, $7<M(g^{'})\leq8$, and
$8<M(g^{'})\leq9$. The scale height for thick disk lies within 806-970 pc 
and the range of the local space density relative to the thin disk for 
the corresponding absolute magnitude interval is (5.25-9.77)\%, in 
agreement with the literature. The halo population is dominant in the 
intervals $4<M(g^{'})\leq5$ , $5<M(g^{'})\leq6$, $6<M(g^{'})\leq7$, and
$7<M(g^{'})\leq8$. The local space density and the axial ratio for the 
brightest interval are rather different than the ones in the other 
intervals and those cited up to now, this is probably due to less stars 
at large distances (see Table 6 and Table 9). For the other absolute magnitude 
intervals the relative local space density, ranging from 0.02\% to 
\%0.13, and the axial ratio with range 0.57-0.78 are also in agreement 
with the literature. The agreement of the luminosity function, derived 
by the combination of the local space densities for  three populations, 
with the Hipparcos one is confirmation for the Galactic model 
parameters.        

\subsection {Parameters determined by means of the procedure in situ}

Let us reiterate the parameters derived by means of stars with 
$5<M(g^{'})\leq10$. The scale-heights for thin and thick disks are 
275 pc and 851 pc respectively, the axial ratio for halo is 0.67, and 
the local space densities for thick disk and halo, relative to the thin 
disk, are 8.32\% and 0.07\% respectively. These values are close to at 
least one, but not to all, of the values estimated for different 
absolute magnitude intervals by means of new approach (see Section 7.1 
and Table 7). For example, the scale height 275 pc mentioned here is 
comparable with the scale height for thin disk estimated for the 
interval $8<M(g^{'})\leq9$, and all the parameters estimated via the 
procedure in situ are close to the corresponding ones estimated for 
the interval $8<M(g^{'})\leq9$ by means of the new procedure. Also, 
these values are in agreement with the parameters cited in the literature.

Although the errors are larger, the luminosity function resulting 
from the comparison of the observed space density functions with the 
Galactic model with the parameters cited above agree with 
the luminosity function of Hipparcos (Jahreiss \& Wielen\ 1997), 
confirming the parameters. 

\begin{table}
\center
\caption{The most appropriate Galactic model parameters (symbols as 
given in the previous tables).}   
{\scriptsize
\begin{tabular}{cccccccc}
\hline
& \multicolumn{2}{c}{Thin disk} & \multicolumn{2}{c}{Thick disk} 
& \multicolumn{2}{c}{Halo} &\\
\hline
$M(g^{'})$ & H (pc)& $n^{*}_{1}$ & H (pc) & $n_{2}/n_{1}(\%)$ & 
$\kappa$ &  $n_{3}/n_{1}(\%)$ & $\odot$ \\
\hline
(5-6]  & 335 & 7.4 & 875 & 9.5 & 0.6 & 0.15 & 7.43 \\
(6-7]  & 325 & 7.4 & 895 & 9.8 & 0.7 & 0.05 & 7.47 \\
(7-8]  & 310 & 7.5 & 805 & 6.5 & 0.8 & 0.02 & 7.48 \\
(8-9]  & 290 & 7.5 & 970 & 5.2 &     &      & 7.52 \\
(9-10] & 265 & 7.6 &     &     &     &      & 7.63 \\
(10-11]& 495 & 8.0 &     &     &     &      & 7.78 \\
(11-12]& 310 & 8.6 &     &     &     &      & 7.92 \\
(12-13]& 275 & 8.1 &     &     &     &      & 8.05 \\
\hline
\end{tabular}  
}
\end{table}

\subsection {How can we decide on the most appropriate Galactic model 
parameters?}

We estimated different sets of Galactic model parameters in two different 
ways, which both are in agreement with the ones appeared so far but differ 
from one another, when a specific one is considered. Now the question is: 
can we select the most appropriate model parameters for our Galaxy? 
We emphasize that the model parameters are absolute - and 
hence mass - dependent. Also, absolutely faint stars do not contribute to 
thick disk and halo populations in the solar neighbourhood. Any procedure 
not regarding these two arguments will result in the estimation of model 
parameters with large range and very likely will differ with the previously 
cited ones, depending on the absolute magnitude interval. 

The dependence of the model parameters on absolute magnitude can be explained 
as such; from the astrophysical point of view, the mass of a star and also its 
chemical composition is two of the fundamental parameters during its formation. 
Hence there is a good correlation between the mass of a main sequence star and 
its absolute magnitude, this in turn is also reflected in its spectral type. 
Therefore, dependence of a model parameters on absolute magnitude is expected. 
Particularly, if we consider the local space density of a thick disk population 
within the local volume, same number of stars with different magnitudes are not 
expected to exist. According to Fig. 8, the thick disk is mostly dominant by the 
stars in absolute magnitude interval of $6<M(g^{'})\leq7$. If for different 
magnitude intervals, the same density law is used, then we can expect to get the 
highest relative local space density within the thick disk for the above given 
interval. Actually, the last column of Table 8 collaborates our expectation, 
where the highest relative local space density, which is 9.77\% and the lowest 
relative local space density, which is 5.25\%, correspond to absolute magnitude 
interval of $6<M(g^{'})\leq7$. The same holds for other model parameters when, 
for instance scale height is considered. Let us consider the thin disk. This 
population is dominant in faint magnitudes. For instance, since at late spectral 
types where stellar masses for the main-sequence stars are relatively small and 
as they occupy the space close to the Galactic plane, one would expect short 
scale heights for these absolute magnitude intervals, whereas one expects large 
scale height for the ones corresponding to relatively bright absolute magnitude 
intervals. This is also the case in our work and is given in Table 7. When local 
space density for thin and thick disks, and for halo or for their both 
combination is considered, dependence of model parameters on absolute magnitude 
would be a result of the luminosity functions adopted up to now for individual 
populations. 

Some researchers restrict their work related with the Galactic model 
estimation with absolute magnitude. The recent work of Robin et al.\ (2003) who 
treated stars with $M(V)\leq8$ can be given as an example. This is in confirmity 
with our work. Finally, we argue that the most appropriate Galactic model 
parameters are those estimated for individual absolute magnitude intervals 
by comparison of the observed space density functions with the corresponding 
(and unique) density law. Our results obtained by this procedure are given 
in Table 16.             

\section*{Acknowledgments}
We wish to thank all those who participated in observations of the field SA
114. The data were obtained through the Isaac Newton Group's Wide Field
Camera Survey Programme, where the Isaac Newton Telescope is operated on the
island of La Palma by the Isaac Newton Group in the Spanish Observatorio del
Roque de los Muchashos of the Instituto de Astrofisica de Canaries. We also
thank CASU for their data reduction and astrometric calibrations, and 
particularly to Professor G. Gilmore for providing the data.

\end{document}